\documentclass[manuscript]{aastex}

\usepackage{graphicx}
\usepackage{amssymb,amsmath}
\usepackage[english]{babel}
\usepackage{url}
\usepackage{ulem}
\usepackage{xcolor}

\shorttitle{Simulations of the ISH fluxes measured by IBEX-Lo}
\shortauthors{Katushkina et al.}

\begin{document}

\title{Interstellar hydrogen fluxes measured by IBEX-Lo in 2009: \\
 numerical modeling and comparison with the data}

\author{O. A. Katushkina\altaffilmark{1}, V. V. Izmodenov\altaffilmark{2,3} and D. B. Alexashov\altaffilmark{3}}
\affil{Space Research Institute of Russian Academy of Sciences, Moscow, 117997, Russia}
\email{okat@iki.rssi.ru}

\author{N. A. Schwadron\altaffilmark{4}}
\affil{University of New Hampshire, Durham, NH 03824, USA}

\and

\author{D. J. McComas\altaffilmark{5}}
\affil{Southwest Research Institute, San Antonio, TX 78228, USA}

\altaffiltext{1}{LATMOS-OVSQ, Guyancourt, France}
\altaffiltext{2}{Lomonosov Moscow State University, Moscow, Russia}
\altaffiltext{3}{Institute of Problems in Mechanics of Russian Academy of Sciences, Moscow, Russia}
\altaffiltext{4}{Southwest Research Institute, San Antonio, TX 78228, USA}
\altaffiltext{5}{University of Texas, San Antonio, TX 78228, USA}

\begin{abstract}
In this paper, we perform numerical modeling of the interstellar hydrogen fluxes measured by IBEX-Lo during orbit 23 (spring 2009) using a state-of-the-art kinetic model of the interstellar neutral hydrogen distribution in the heliosphere. This model takes into account
 the temporal and heliolatitudinal variations of the solar parameters as well as non-Maxwellian kinetic
 properties of the hydrogen distribution due to charge exchange in the heliospheric interface.

We found that there is a qualitative difference between the IBEX-Lo data and the modeling results obtained with the three-dimensional, time-dependent model.
 Namely, the model predicts a larger count rate
 in energy bin~2 (20-41~eV) than in energy bin~1 (11-21~eV), while the data shows the opposite case.

 We perform study of the model parameter effects on the IBEX-Lo fluxes
 and the ratio of fluxes in two energy channels. We shown that the most important parameter, which has a major influence on the ratio
 of the fluxes in the two energy bins, is the solar radiation pressure. The parameter fitting procedure shows that the best agreement between the model result and the data occurs in the case when the ratio of the solar radiation pressure to the solar gravitation, $\mu_0$, is 1.26$^{+0.06}_{-0.076}$, and the total ionization rate of hydrogen at 1~AU is $\beta_{E,0}=3.7^{+0.39}_{-0.35}\times 10^{-7}$~s$^{-1}$.
We have found that the value of $\mu_0$ is much larger than $\mu_0=0.89$, which is the value derived from the integrated solar Lyman-alpha flux data for the period of time studied. We discuss possible reasons for the differences.

\end{abstract}


\section{Introduction: a brief historical review}

The first evidence for the presence of the interstellar hydrogen atoms (H atoms) in the interplanetary medium
was obtained in the late 1950s from the night-flight rocket measurements of the diffuse UV
emission at the H Lyman-$\alpha$ line with a central wavelength of 1215.6~{\AA} \citep{kupperian_etal_1959, shklovsky_1959}. It was suggested that the observed emission was caused by either the scattering of the solar Lyman-$\alpha$ radiation by hydrogen atoms in the interplanetary medium or the galactic Lyman-$\alpha$ emission. Additional more detailed experiments performed on board the OGO-5 satellite in 1969-1970
\citep[see][]{bertaux_blamont_1971, thomas_krassa_1971}
provided maps of the Lyman-$\alpha$ intensities and showed the 50$^{\circ}$ apparent displacement of the maximum
emissivity region (MER) between the measurements from 1969 September to 1970 April.
This displacement was explained by the parallax-effect caused by Earth's motion around the Sun and was proof that the source of the measured Lyman-$\alpha$
emission is located at several (2-3) AU from the Sun. \citet{bertaux_blamont_1971} and \citet{blum_fahr_1972} have interpreted these
observations in terms of the neutral ``interstellar wind''. Namely, the neutral interstellar H atoms (ISH) penetrate to the heliosphere due to relative motion of the Sun through the
local interstellar medium (LISM). Inside the heliosphere they
scatter the solar Lyman-$\alpha$ photons. As a result, backscattered radiation is formed and can be measured,
e.g., at Earth's orbit.

Measurements of the backscattered solar Lyman-$\alpha$ radiation in the heliosphere stimulated the development of theoretical models that
 describe the propagation of the interstellar H atoms from the LISM to the vicinity of the Sun.
The first generation of the models is so-called the ``cold model'' proposed by \citet{fahr_1968} and \citet{blum_fahr_1970}.
They assumed that the LISM is cold, i.e. all of the H atoms have the same velocity and
penetrate deeply to the heliosphere due to the relative motion of the Sun through the surrounding interstellar matter. Analytical expressions for the number density of H atoms in the heliosphere and the corresponding
 intensity of the backscattered Lyman-alpha radiation obtained for the cold model are given by \citet{dalaudier_etal_1984} in the ``attractive''case
  (when the solar gravitation attractive force is larger than the solar radiative repulsive force) and by \citet{lallement_etal_1985a} in the
  opposite ``repulsive'' case.
 Measurements of the interplanetary Lyman-$\alpha$
glow using the hydrogen absorption cell onboard Prognoz-5 spacecraft allowed to estimate of the LISM temperature
($\sim$ 8000~K) that is not negligible \citep[see, e.g., ][]{bertaux_etal_1977}. Therefore, a second generation of the
models, so-called ``hot models'', were developed. The hot model takes into account a realistic temperature
and the corresponding thermal velocities of H atoms in the LISM.
\citet{meier_1977} and \citet{wu_judge_1979} have presented an analytical solution for the hot model of the ISH velocity distribution.
 They take into account the solar gravitational attractive force, the solar radiative repulsive force, and losses of H atoms due to photoionization and charge exchange with the solar wind (SW) protons. In the classical hot model, it is assumed that the problem is stationary and axisymmetric, and that the ISH velocity distribution function at infinity (i.e. in the LISM) is uniformly Maxwellian.
The mathematical formulation of the hot model and a review of its results and further modifications can be found in \cite{izmod_issi_2006}.

In the 1980-1990s, the classical hot model was widely used to interpret experimental data for interstellar hydrogen
 in the heliosphere (namely, measurements of backscattered Lyman-alpha radiation and pickup ions). However, it became clear that
 the classical hot model is appropriate for general estimates of the ISH parameters in the heliosphere, but it is not sufficiently accurate for studying
 more detailed effects.

 In general, there are two ways to improve the classical hot model.
The first way is to take into account the temporal and heliolatitudinal variations of the
solar parameters (namely, parameters of the solar radiation and the SW).
Temporal variations are caused by the 11-year cycle of the solar activity and have been considered in many works
\citep[e.g.,][]{bzowski_rucinski_1995, summanen_1996, bzowski_etal_1997, pryor_etal_2003, bzowski_2008}.
The heliolatitudinal variations are connected with the nonisotropic SW structure.
\citet{joselyn_holzer_1975} were the first to show that the nonisotropic SW would
strongly affect the ISH distribution in the heliosphere. The signatures of the heliolatitudinal variations
of the SW were found in the measurements of the Lyman-alpha intensities on board Mariner-10, Prognoz-6, and the Solar and Heliospheric
Observatory (SOHO)/SWAN \citep[see, e.g., review by][]{bertaux_etal_1996} and were later confirmed by direct measurements
by the Ulysses spacecraft out of the ecliptic plane \citep{mccomas_etal_2003, mccomas_etal_2006, mccomas_etal_2008}.
Several authors \citep{lallement_etal_1985b, pryor_etal_2003, nakagawa_etal_2009} assumed some analytical relations
for the heliolatitudinal variations of the hydrogen ionization rate and took them into account in the frame of the hot model.

The second way that the classical hot model can be improved is to take into account disturbances of the ISH flow in the region
of interaction between the SW and the charged component of the LISM (in the literature this region is called the heliospheric interface).
Theoretical study of the SW/LISM interaction began with the pioneering works by \citet{parker_1961} and \citet{bkk_1970}.
In these works, the supersonic fully ionized SW flow interacts with the fully ionized interstellar plasma or with the
interstellar magnetic field (IsMF), but interstellar neutral atoms were not taken into account. By the 1970s \citep{wallis_1975}
 it was realized that the hydrogen atoms interact with protons through charge exchange ($H+H^+\rightleftarrows H^++H$), which leads to
 an interchange of the momentum and energy between the charged and neutral components and dynamically influences the heliospheric interface structure
 \citep{baranov_etal_1981}. The first self-consistent two-component model of the interaction between the supersonic SW flow and the partially
 ionized supersonic interstellar wind was developed by \citet{bm_1993}. In this model, the ideal gasdynamical Euler equations for the charged component
 are solved self-consistently with the kinetic Boltzmann equation for H atoms. Only a kinetic approach is valid for the description of the ISH distribution because the mean free path of H atoms with respect to the charge exchange with protons is comparable to the size of the heliosphere \citep[for a review, see][]{izmod_2001_cospar, izmod_etal_2001}.
   It was shown that in the case of the supersonic interstellar flow, the heliospheric interface consists of four regions
  separated by three discontinuities: the heliopause (HP) is a contact discontinuity distinguishing
 the SW plasma from the interstellar plasma, and the Termination Shock (TS), and the Bow Shock (BS) are the shocks where the SW and the interstellar
 wind, respectively, become subsonic. Note that the Bow shock may be absent in the presence of a strong IsMF that makes
 the interstellar flow subsonic \citep[see, e.g.,][]{izmod_etal_2009, mccomas_etal_2012}.

 The interstellar H atoms penetrate through all of the discontinuities into the heliosphere due to their large mean free path.
 However, the charge exchange with protons leads to significant disturbances of the hydrogen flow in the heliospheric interface. First,
 the heliospheric interface may be considered as a filter for the primary interstellar H atoms \citep{izmod_2007} because only a small fraction of them
 can reach the inner part of the heliosphere.
 Second, new ``secondary'' H atoms are created in the heliospheric interface by charge exchange. These secondary atoms have the individual velocities of their original parent protons.
 Therefore, the velocity distribution function of newly created atoms depends on the local plasma properties, which are different in the various regions of the heliospheric interface. Thus, the mixture of the primary and secondary
 interstellar H atoms penetrates inside the heliosphere and their properties depend on both the LISM parameters and the plasma distribution in the heliospheric interface.
 This also means that the classical specification of the boundary conditions in the hot models as a Maxwellian distribution in the LISM (ignoring the region
 of SW/LISM interaction) is a crude approximation.
 Disturbances of the ISH flow in the heliospheric interface were included in the hot model using the different approaches suggested by \citet{scherer_etal_1999, bzowski_etal_2008, nakagawa_etal_2008, katush_izmod_2010}, and \citet{izmod_etal_2013}.

Since the 1980s, the classical hot model and its advanced modifications have been widely used to interpret the experimental data on the backscattered solar Lyman-$\alpha$ radiation
\citep[see, e.g., ][]{lallement_etal_1985a, costa_etal_1999, bzowski_2003, pryor_etal_2008} and pickup ions \citep[e.g.,][]{bzowski_etal_2008, bzowski_etal_2009}.
For example, the bulk velocity and temperature of the interstellar hydrogen far away from the Sun
(at 80-100~AU) were obtained from theoretical analyses of the experimental data on the
 Lyman-alpha radiation \citep{bertaux_etal_1985, costa_etal_1999}, while the number density of hydrogen at the TS
 was derived from pickup ions measurements by Ulysses/SWICS \citep[see for review][]{bzowski_etal_2009}.
It was shown that the ISH flow in the heliosphere is decelerated and heated compared with the parameters of the pristine interstellar wind. These effects are explained by the presence of the secondary
interstellar atoms, which are created from the interstellar protons near the HP
and have smaller velocity and larger temperature compared with the original interstellar parameters.

Since 2009, fluxes of the ISH were measured in situ for the first time near Earth's orbit by the IBEX-Lo sensor \citep{fuselier_etal_2009, moebius_etal_2009} on board the Interstellar Boundary Explorer (IBEX) spacecraft \citep{mccomas_etal_2009}.
The main goal of the IBEX mission is to study the three-dimensional (3D) structure of the heliospheric boundary through measurements of the energetic
neutral atoms (ENAs) created in the heliospheric interface. Recently, \citet{mccomas_etal_2014} has summarized the
IBEX ENA results obtained over five years of observations. IBEX has two sensors for measurements of the heliospheric and interstellar neutrals (hydrogen, helium,
 oxygen, and neon) with different energies. The IBEX-Hi sensor measures ENAs with energies from $\sim$300 eV to 6 keV \citep{funsten_etal_2009}. The IBEX-Lo sensor
(with energy range $\sim$10~eV to 2~keV) measures ENAs and
low energetic interstellar atoms \citep{fuselier_etal_2014, kubiak_etal_2014, park_etal_2014}. McComas et al. (2015b) have summarized the results
obtained during six years of IBEX-Lo measurements of low energetic interstellar neutrals.

IBEX-Lo data on the ISH fluxes are an effective tool
for verifying of the theoretical models of the ISH distribution and can be used to fit the model parameters and improve
our knowledge of the LISM and the heliospheric interface structure.
Previously, \citet{saul_etal_2012, saul_etal_2013} presented the IBEX-Lo hydrogen data obtained during spring passage in 2009-2012 and showed
that the signal strongly decreased with time and almost disappeared in 2012 (most probably due to the arising of the solar radiation and ionization after the solar minimum in 2009).
\citet{schwadron_etal_2013} presented an analysis of the 2009-2011 data using the hot model without considering the time-dependent
effects and influence of the heliospheric interface. Through a comparison between the hot model and IBEX data, \citet{schwadron_etal_2013} found the best-fit model parameters
(these parameters include solar radiation pressure, velocity, and temperature of the ISH beyond the TS).
  Further investigations showed
that the procedure of response-function integration of H fluxes in \citet{schwadron_etal_2013}
was not well resolved. We have developed a more accurate response-function integration in our work.

The goal of this paper is to apply the state-of-the-art 3D time-dependent kinetic model of the ISH distribution developed by \citet{izmod_etal_2013}
to simulations
and analysis of the ISH fluxes measured by IBEX-Lo during the spring passage in 2009 (namely, orbit 23 when the largest fluxes were measured).
In section~\ref{section_model}, the mathematical description of the model and its input parameters are provided. Section~3 briefly describes the IBEX-Lo
ISH data.
In section~\ref{res}, we compare the results of the state-of-the-art numerical model with the data and investigate
the influence of some model parameters on the ISH fluxes.
Section~5 presents the results of the stationary version of the model. We perform a parametric study of the different magnitudes of the hydrogen ionization rate
and the solar radiation pressure.
We shown that the model parameter $\mu$ (which characterizes the ratio between solar radiation pressure and gravitation) is critically important for the ISH fluxes measured by IBEX-Lo.
Small variations of $\mu$ lead to
significant changes in the ratio of counts in the IBEX-Lo energy bins~1 and 2. Therefore, precise knowledge of the
solar Lyman-alpha flux at the line center (which determines the magnitude of $\mu$ for zero radial atom's velocity) and the shape of the
Lyman-alpha spectrum (corresponding to the velocity dependence of $\mu$) is necessary for analysis of the IBEX-Lo ISH data.
In section~\ref{section_fitting}, we perform a fitting of the IBEX-Lo data for orbit~23 to estimate the solar parameters which allow us to obtain agreement with IBEX data in the frame of the stationary model. We obtained magnitude of $\mu$ that is considerably
larger than that derived from direct measurements of the solar radiation. This raises questions about our current understanding of the hydrogen distribution
near the Sun, as well as for absolute calibration of the solar Lyman-alpha flux data and accuracy of the IBEX-Lo instrumental response. These aspects are discussed in section~7.

This study is part of a coordinated set of papers on interstellar neutrals as measured by IBEX. \citet{mccomas_etal_2015_overview} provide an overview of this Astrophysical Journal Supplement Series Special Issue.


\section{Model of the ISH distribution}\label{section_model}

In this section, we briefly describe the advanced kinetic model of the ISH distribution in the heliosphere.
This model was proposed by \citet{izmod_etal_2013} and previously applied for the analysis of Lyman-$\alpha$ data in \citet{katushkina_etal_2013, katushkina_etal_mnras_2015}. The model is a 3D time-dependent version of the classical hot model with specific boundary conditions at 90~AU
based on the results of a global self-consistent model of the heliospheric interface. Below, we will refer to this model as the base model.
The outer boundary of the computational region is set at 90~AU from the Sun.

We only consider the interstellar fraction of H atoms in the heliosphere, which is a mixture of the primary and secondary
interstellar atoms. The secondary atoms are created by charge exchange between the primary atoms and
the interstellar protons outside the HP. We do not consider the heliospheric atoms created through charge exchange
with the SW protons and pickup ions inside the heliosphere because they have large energy and do not contribute to the low energetic interstellar fraction that we are
interested in here. Therefore, charge exchange and photoionization inside the heliosphere lead to the loss of interstellar H atoms.

The distribution of interstellar H atoms is described by a kinetic equation:
\begin{eqnarray}
   \frac{\partial f(\textbf{r},\textbf{w},t)}{\partial t}
   + \textbf{w}\cdot\frac{\partial f (\textbf{r},\textbf{w},t)}{\partial \textbf{r}}+ & \nonumber \\
   \frac{\textbf{F}(\textbf{r},t,\lambda,w_r)}{m_H}\cdot \frac{\partial f(\textbf{r},\textbf{w},t)}{\partial \textbf{w}} &
   = -\beta(r,t,\lambda)\cdot f(\textbf{r},\textbf{w},t).\label{Boltzmann}
\end{eqnarray}
Here, $f(\textbf{r},\textbf{w},t)$ is the
velocity distribution function of H atoms, $\textbf{w}$ is the individual velocity of an H atom, and $m_H$ is the mass of an H atom.
$\textbf{F}$ is a force acting on each atom in the heliosphere. This force is a sum of the solar gravitational attractive force ($\textbf{F}_g$) and the solar radiative repulsive force ($\textbf{F}_{rad}$).
Both forces are proportional to $\propto 1/r^2$ ($r$ is the heliocentric distance), and therefore it is convenient to introduce the dimensionless parameter $\mu=|\textbf{F}_{rad}|/|\textbf{F}_g|$. Then,
\[
\textbf{F}=\textbf{F}_g+\textbf{F}_{rad}=(1-\mu (t, \lambda ,w_r))\textbf{F}_{g}=-m_H \frac{(1-\mu (t, \lambda ,w_r)GM_{s}}{r^{2}}\cdot \frac{\textbf{r}}{r},
\]
where $G$ is the gravitational constant and $M_{s}$ is the mass of the Sun.
In general, the parameter $\mu$ depends on the time
($t$), heliolatitude ($\lambda$), and the radial component of the atom's velocity ($w_{r}$).

The right-hand side of equation~(\ref{Boltzmann}) represents the loss of atoms due to ionization processes, namely, charge exchange
($H + H^{+} = H^{+} + H$) and photoionization ($H+h\nu=H^{+}+e$).
Electron impact ionization is not taken into account because, as was shown by \citet{bzowski_fondue}, the rate of
electron impact ionization is at least one order of magnitude smaller than the total hydrogen ionization rate at 1~AU
from the Sun.
The coefficient $\beta(r,t,\lambda)$ is the effective ionization rate:
$\beta(r,t,\lambda)=\beta_{ex}(r,t,\lambda)+\beta_{ph}(r,t,\lambda)$, where $\beta_{ex}$ and $\beta_{ph}$
are the rates of charge exchange and photoionization, respectively. These rates decrease with distance from the Sun as $\sim1/r^{2}$,
since these values are proportional to the number density of the SW protons and flux of the solar EUV photons.
Therefore,
  \[
 \beta(r,t,\lambda)=\left( \beta_{ex,E}(t,\lambda)+\beta_{ph,E}(t,\lambda) \right)
 \left(\frac{r_{E}}{r}\right)^{2} = \beta_{E}(t,\lambda)\left(\frac{r_{E}}{r}\right)^{2},
  \]
where $r_{E}=1$~AU, subscript $E$ indicates that the values are taken at 1 AU. Ionization rates depend on time and heliolatitude due to the temporal and latitudinal variations of the SW mass flux and solar EUV radiation.
The functions $\mu(t,\lambda,w_r)$, $\beta_{ex,E}(t,\lambda)$, and $\beta_{ph,E}(t,\lambda)$ adopted in our model
are obtained from different experimental data. Detailed descriptions of these functions will be given below in this section.

Kinetic equation~(\ref{Boltzmann}) is a linear partial differential equation that can be solved by the method of characteristics. The solution of this equation is as follows:
\begin{equation*}
   f(\textbf{r},\textbf{w},t)=
  f_b(\textbf{r}_{0},\textbf{w}_{0})\,
  \exp\left( -\int_{t_0}^{t} \beta(r,t,\lambda)dt\right),
\end{equation*}
where $f_{b}(\textbf{r}_{0}, \textbf{w}_{0})$ is the velocity distribution function of hydrogen
atoms at the outer boundary (determined by the stationary boundary conditions at 90~AU);
$\textbf{r}_{0},\textbf{w}_{0}, t_{0}$ are the position, velocity, and time when the atom crossed the outer boundary and
entered to the computational region. The integration in the last equation is performed along the atom's trajectory.

Charge exchange in the heliospheric interface leads to disturbances of the ISH flow and, as a result, the velocity distribution functions of the primary and secondary interstellar atoms inside the HP are not Maxwellian \citep{izmod_etal_2001}. A detailed description of the non-Maxwellian properties of the hydrogen distribution
at 90~AU is presented by \citet{izmod_etal_2013}.
 Therefore, the specific non-Maxwellian boundary conditions at 90~AU are necessary to take into account the influence of the heliospheric interface. \citet{katush_izmod_2010, katush_izmod_2012} discussed several kinds of the boundary velocity
distribution function based on the results of the self-consistent axisymmetrical kinetic-gasdynamic model of the SW/LISM interaction \citep{bm_1993}.
For the present work, the boundary conditions in the form of a 3D normal distribution were adopted at 90~AU separately for the primary and secondary interstellar atoms.
This form of the boundary conditions allows us to include all zero, first, and second moments of the velocity distribution function. In the 3D case without any symmetries, the analytical expression for the
 adopted boundary distribution function is as follows:
\begin{eqnarray}
f_b(\textbf{r}_0,\textbf{w}_0)&=\left(\frac{m_H}{2\pi k}\right)^{3/2}\frac{n_H}{\sqrt{D}}\cdot \exp(-\frac{m_H}{2D \cdot k}(C_{22}(V_{\rho}-w_{\rho,0})^2+ \\
  &+ C_{33}(V_{\varphi}-w_{\varphi,0})^2+C_{11}(V_{z}-w_{z,0})^2 + \nonumber\\
  &+ 2C_{12}(V_{\rho}-w_{\rho,0})(V_z-w_{z,0})+2C_{13}(V_z-w_{z,0})(V_{\varphi}-w_{\varphi,0})+ \nonumber\\
  &+ 2C_{23}(V_{\rho}-w_{\rho,0})(V_{\varphi}-w_{\varphi,0}))) \nonumber,  \label{3Dnorm}
\end{eqnarray}
where
\begin{eqnarray*}
  D=T_zT_{\rho}T_{\varphi}+2T_{z\rho}T_{\varphi\rho}T_{\varphi z}-T_{\rho}T_{\varphi z}^{2}-T_{z}T_{\varphi \rho}^{2}-T_{\varphi}T_{z\rho}^{2}, \\
  C_{11}=T_{\varphi}T_{\rho}-T_{\varphi \rho}^2; C_{22}=T_{\varphi}T_{z}-T_{\varphi z}^2; C_{33}=T_{z}T_{\rho}-T_{z \rho}^2; \\
  C_{12}=T_{\varphi z}T_{\varphi \rho}-T_{\varphi}T_{z \rho}; C_{13}=T_{\varphi z}T_{\rho}-T_{\varphi \rho}T_{z \rho}; C_{23}=T_{z \rho}T_{\varphi z}-T_{z}T_{\varphi \rho}.
\end{eqnarray*}
Here, $n_H$ is number density of the atoms, $(V_{\rho}, V_{\varphi}, V_z)$ are components of the bulk atom's velocity in a cylindrical system of coordinates
(where the axis $\textbf{e}_z$ is opposite to the direction of the interstellar wind flow relative to the Sun, and the axes $\textbf{e}_{\rho}$ and $\textbf{e}_{\varphi}$
are linear and orthogonal and make a right-handed orthogonal system of coordinates), $(T_{\rho}, T_{\varphi}, T_z)$ are the kinetic ``temperatures'' of H atoms,
and $(T_{\varphi \rho}, T_{\varphi z}, T_{z \rho})$ are correlation coefficients. Namely,
\begin{eqnarray*}
 n_H(\textbf{r}_0) &=\int f(\textbf{r}_0,\textbf{w}) \, d\textbf{w} \\
 V_i(\textbf{r}_0) &=\left( \int f(\textbf{r}_0,\textbf{w}) \cdot w_i \, d\textbf{w} \right) / n_H(\textbf{r}_0) \\
 T_i(\textbf{r}_0) &\sim \left( \int f(\textbf{r}_0,\textbf{w})\cdot (w_i-V_i)^2 \, d\textbf{w}\right) / n_H(\textbf{r}_0) \\
 T_{i j}(\textbf{r}_0) &\sim \left( \int f(\textbf{r}_0,\textbf{w})\cdot (w_i-V_i)(w_j-V_j) \, d\textbf{w} \right) / n_H(\textbf{r}_0).
\end{eqnarray*}
All of these parameters ($n_H, \textbf{V}, T_i, T_{i j}$) depend on the position at the boundary sphere (i.e. on two spherical angles) and are taken from results of
the new self-consistent kinetic-MHD model of the heliospheric interface recently developed by our Moscow group.
This model is a sophisticated 3D stationary version of the original model of \citet{bm_1993} with the kinetic description of H atoms. It takes into account
the heliospheric and interstellar magnetic fields and heliolatitudinal dependence of the SW parameters at 1~AU.
This model and its results are described in detail in a companion paper \citet{izmod_alexash_2015} in this Special Issue. The following
LISM parameters are used: the number density of protons is $n_{p,LISM}=0.04$~cm$^{-3}$, the number density of H atoms is
$n_{H,LISM}=0.14$~cm$^{-3}$, the velocity of the interstellar wind is $V_{LISM}$=26.4~km/s and its direction is taken from the Ulysses interstellar neutral He data analysis reported by \citet{witte_2004} i.e. the ecliptic (J2000) longitude is 75.4$^{\circ}$ and the latitude is -5.2$^{\circ}$,
LISM temperature is $T_{LISM}$=6530~K, IsMF is $B_{LISM}$=4.4~$\mu G$, the angle between $\textbf{B}_{LISM}$ and $\textbf{V}_{LISM}$ is 20$^{\circ}$,
and the $(\textbf{B},\textbf{V})_{LISM}$-plane coincides with the Hydrogen Deflection Plane (HDP)
first proposed by \citet{lallement_etal_2005} and then slightly changed in \citet{lallement_etal_2010}. In ecliptic (J2000) coordinates,
the vector $\textbf{B}_{LISM}$ has longitude 62.49$^{\circ}$ and latitude -20.79$^{\circ}$.

Thus, the procedure to obtain the ISH velocity distribution function inside the heliosphere consists of two
consecutive steps:
\begin{enumerate}
 \item in the first step, the parameters of the the primary and secondary interstellar atoms at a sphere with a radius of 90~AU are obtained from
 the global 3D stationary kinetic-MHD model of the SW/LISM interaction; and
  \item in the second step, the kinetic equation~(\ref{Boltzmann}) is solved with the boundary conditions~(2)
  separately for the primary and secondary interstellar atoms. The total velocity distribution function is the sum of the distribution functions of the primary and secondary atoms.
\end{enumerate}
This procedure allows us to take into account simultaneously the local temporal and heliolatitudinal variations of the
SW and solar radiation (which are extremely important for the ISH parameters at small heliocentric distances) and
the global effects of the charge exchange in the heliospheric interface (which lead to the non-Maxwellian features of the hydrogen velocity distribution function far away from the Sun).

Below, we describe the model parameters $\mu$ and $\beta_E$ based on different experimental data and several assumptions.

\subsection{Parameter $\mu(t,\lambda,w_r)$}

\begin{figure}
\includegraphics[scale=0.8]{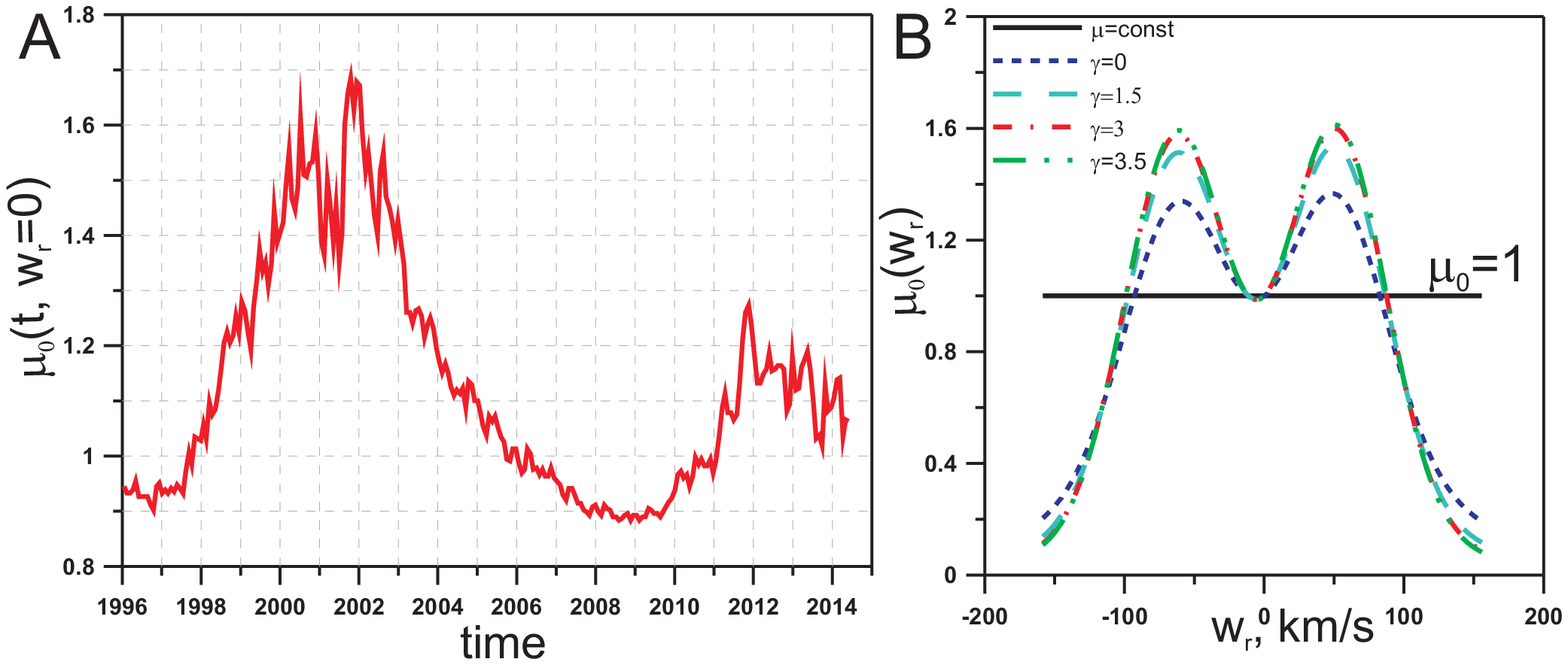}
\caption{A. Temporal variations of parameter $\mu_0$ (for zero radial atom's velocity and heliolatitude). It is calculated from
the integrated solar Lyman-alpha flux (known from LASP database) and the formula of \citet{emerich_etal_2005} for transformation of
the integrated flux to the flux at line center. B. Velocity dependence of $\mu_0(w_r)$ for different $\gamma$ (see formula~3).
}\label{mu_t_vr}
\end{figure}

The parameter $\mu_{0}$ at zero heliolatitude ($\lambda=0$) and a zero radial atom's velocity ($w_r=0$) can be calculated
from the total solar line-integrated Lyman-alpha flux ($F_\mathrm{solar}(t)$) by the following equation:
\[
\mu_{0}(t)=0.64\cdot 10^{11}\cdot (F_\mathrm{solar}(t)\cdot 10^{-11})^{1.21}/F_\mathrm{solar,0},
\]
 where $F_\mathrm{solar,0}=3.32\cdot 10^{11} \, ph/(s \, cm^2 \, \mathrm{{\AA}})$.
This expression for the transformation of the total solar Lyman-alpha flux to the flux at the line center is found by \citet{emerich_etal_2005}.
Note that we previously \citep[e.g. in][]{izmod_etal_2013} used a simplified relation (just a factor of 0.9) for this transformation that is not correct during solar minima.
The total solar Lyman-$\alpha$ flux ($F_{solar}(t)$) is taken
 from the LASP Interactive Solar IRradiance Data center (\url{http://lasp.colorado.edu/lisird/lya/}). From this database, we obtain the solar Lyman-$\alpha$ flux as a function of time with a resolution of one day. These data are then adjusted to 1 AU from Earth's orbit and averaged over one Carrington rotation (about 27 days).
 Temporal variations of $\mu_0$ are presented in Fig.~\ref{mu_t_vr}~A.

The original solar Lyman-alpha profile that determines the velocity dependence of the solar radiation pressure was measured by the SUMER spectrometer
 on board SOHO \citep[see, e.g.,][]{lemaire_etal_2005}.
To take into account the dependence of $\mu$ on a radial atom's velocity ($w_r$), an analytical expression proposed by \citet{bzowski_2008}
and generalized by \citet{schwadron_etal_2013} is used in our model. Namely,
\[
\mu_{0}(t,w_r)=\mu_{0}(t) \cdot F_{SB}(w_r),
\]
where
\begin{equation}
F_{SB}(w_r)=exp(-C\,w_r^2) \cdot \frac {[1+(1+\gamma) (D \, exp(F \, w_r-G \, w_r^2)+H \, exp(-P \, w_r-Q \, w_r^2))] } {[1+(D+H) (1+\gamma)]}, \label{mu_vr}
\end{equation}
where the constants are the following: $C=3.8312\cdot 10^{-5}$,
$D=0.73879$, $F=4.0396\cdot 10^{-2}$, $G=3.5135\cdot 10^{-4}$, $H=0.47817$, $P=4.6841\cdot 10^{-2}$, and
$Q=3.3373\cdot 10^{-4}$ \citep[see,][]{bzowski_2008}. If $w_r=0$, then $F_{SB}=1$ and $\mu=\mu_0$. $\gamma$ characterizes the wings of the velocity-dependent profile
(larger $\gamma$ corresponds to larger wings, but this dependence is very weak for $\gamma >1.5$, see Fig.~\ref{mu_t_vr}~B).
 Changes in $\gamma$ influence those atoms with individual velocities of $~$30-70~km/s.
 By default, we perform calculations with $\gamma=0$. This case corresponds to the original expression from \citet{bzowski_2008}, and we indicate
  specifically if other values of $\gamma$ are used.

To determine the heliolatitudinal dependence of $\mu$, we use the following expression from \citet{pryor_etal_1992}:
\[
 \mu(t,\lambda,w_r)=\mu_\mathrm{pole}(t,w_r)+cos^{2}(\lambda)\cdot (\mu_{0}(t,w_r)-\mu_\mathrm{pole}(t,w_r)),
\]
where
\[
\mu_\mathrm{pole}(t,w_r)=\frac{0.64\cdot 10^{11}((F_\mathrm{solar}(t)-\Delta F_\mathrm{solar})\cdot 10^{-11})^{1.21}}{F_\mathrm{solar,0}} \cdot F_{SB}(w_r)
\]
and $\Delta F_\mathrm{solar}\approx 0.05\cdot 10^{11}$ ph/(cm$^2$s), this value is taken from \citet{pryor_etal_1998}.
The heliolatitudinal dependence of $\mu$ is quite weak (variations of $\mu$ are not more than 0.1).

\subsection{Parameter $\beta_{E}(t,\lambda)$}

Temporal variations of the photoionization and charge exchange ionization rates in the ecliptic plane are obtained based on
the SOLAR2000 and OMNI2 databases. The data are averaged over one Carrington rotation of the Sun in order to exclude any possible longitudinal variations.
Therefore, the time resolution in our model is about 27 days. The heliolatitudinal variations of the ionization rate are adopted from the results of
an analysis of the full sky-maps in the backscattered Lyman-alpha intensities measured by SOHO/SWAN \citep{quemerais_etal_2006_jgr, lallement_etal_2010}.
A detailed description of the adopted ionization rates can be found in \citet{izmod_etal_2013}.

Note that an alternative method for the reconstruction of the heliolatitudinal variations of the SW parameters has been proposed by \citet{sokol_etal_2013}.
Their method is based on deriving the SW speed profile (over latitude) from interplanetary scintillation data, direct measurements of Ulysses
during its fast latitudinal scans, and assuming a linear correlation between the speed and density of the SW.
 However, \citet{katushkina_etal_2013} have shown that the results of \citet{sokol_etal_2013} are inconsistent with the Lyman-alpha intensity maps measured by
 SOHO/SWAN during the maximum of the solar activity (most likely due to an incorrect assumption on the linear correlation between the SW speed and density,
 which does not work at the solar maximum). At the same time, during the solar minima conditions (considered here), both models provide
 qualitatively the same heliolatitudinal dependence of the SW mass flux.

\section{Measurements of the ISH fluxes by IBEX-Lo}

IBEX is a spinning spacecraft with the spin-axis reoriented toward the Sun at each orbit or orbit arc. The direction of the spin-axis remains fixed between
each re-orientation maneuver. Each orbit around the Earth takes approximately 7-9 days.
In our simulations, we use actual trajectory, velocities, and spin-axis orientations of IBEX, which are available at the webpage of the IBEX public Data Release~6 (\url{http://ibex.swri.edu/ibexpublicdata/Data_Release_6/}).
Simulations are performed for orbit 23, corresponding to the dates from 2009 March 27 to April 2.
We choose this orbit because \citet{schwadron_etal_2013} has shown that it corresponds to a peak of the ISH fluxes as measured by IBEX-Lo
(this means that the signal-to-noise ratio should be the largest for this orbit). Also, the data taken within this orbit
are not contaminated by the Earth's magnetosphere and the background is at a low level
(i.e. it is a ``good time'' for observations of the ISH).
In our simulations, we use the actual time periods of observations listed in Table~1 of \citet{schwadron_etal_2013}.
We consider only the first two IBEX-Lo energy channels (bin~1: 11-21~eV and bin2: 20-41~eV), because the most of the low energetic interstellar H
atoms should appear in these channels.

\begin{figure}
\includegraphics[scale=0.8]{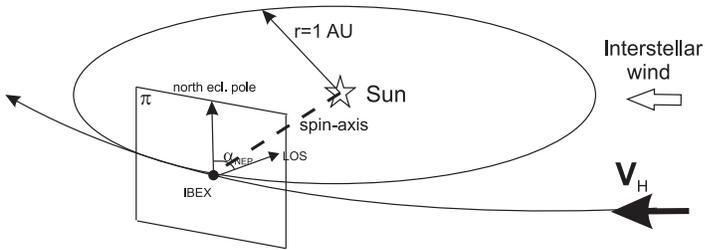}
\caption{Schematic representation of IBEX's observational geometry. Plane $\pi$ is the plane of measurements
that is perpendicular to the spacecraft-Sun vector. Angle $\alpha_{NEP}$ is counted in plane $\pi$ from the north
ecliptic pole and characterizes the direction of the line of sight.
}\label{ibex_geom}
\end{figure}

IBEX measures the fluxes of the interstellar neutrals in the plane perpendicular to the spin-axis (plane $\pi$ in Fig.~\ref{ibex_geom}).
 The line of sight in this plane can be described by the angle from the direction of the north ecliptic pole (NEP angle
or $\alpha_{NEP}$). The IBEX-Lo sensor has a collimator with a 7$^{\circ}$ FWHM.

The IBEX-Lo hydrogen data processed and presented by \citet{schwadron_etal_2013} are averaged over the ``good'' times of
observations during each orbit. To be consistent with the data, we calculate the ISH fluxes as function of NEP angle for each good day during orbit 23 and then average the results over all of the days. Calculations are performed for the lines of sight characterized by $\alpha_{NEP}\in[60^{\circ}, 114^{\circ}]$ with steps of 1$^{\circ}$. Then, the obtained fluxes
 are accumulated for each 6$^{\circ}$ bin with $\Delta \alpha=6^{\circ}$ (in the same way as it is done for the IBEX data).
 For comparison with the real IBEX-Lo data, one must convert the fluxes calculated in the model to count rate (number of counts per second).
 This technical procedure is described in Appendix~A.

\section{Results of the time-dependent model}\label{res}

In this section, we present the results of calculations of the ISH fluxes for IBEX-Lo energy bins 1 and 2 performed in the frame of the time-dependent model described
in section~\ref{section_model}.

\begin{figure}
\includegraphics[scale=0.8]{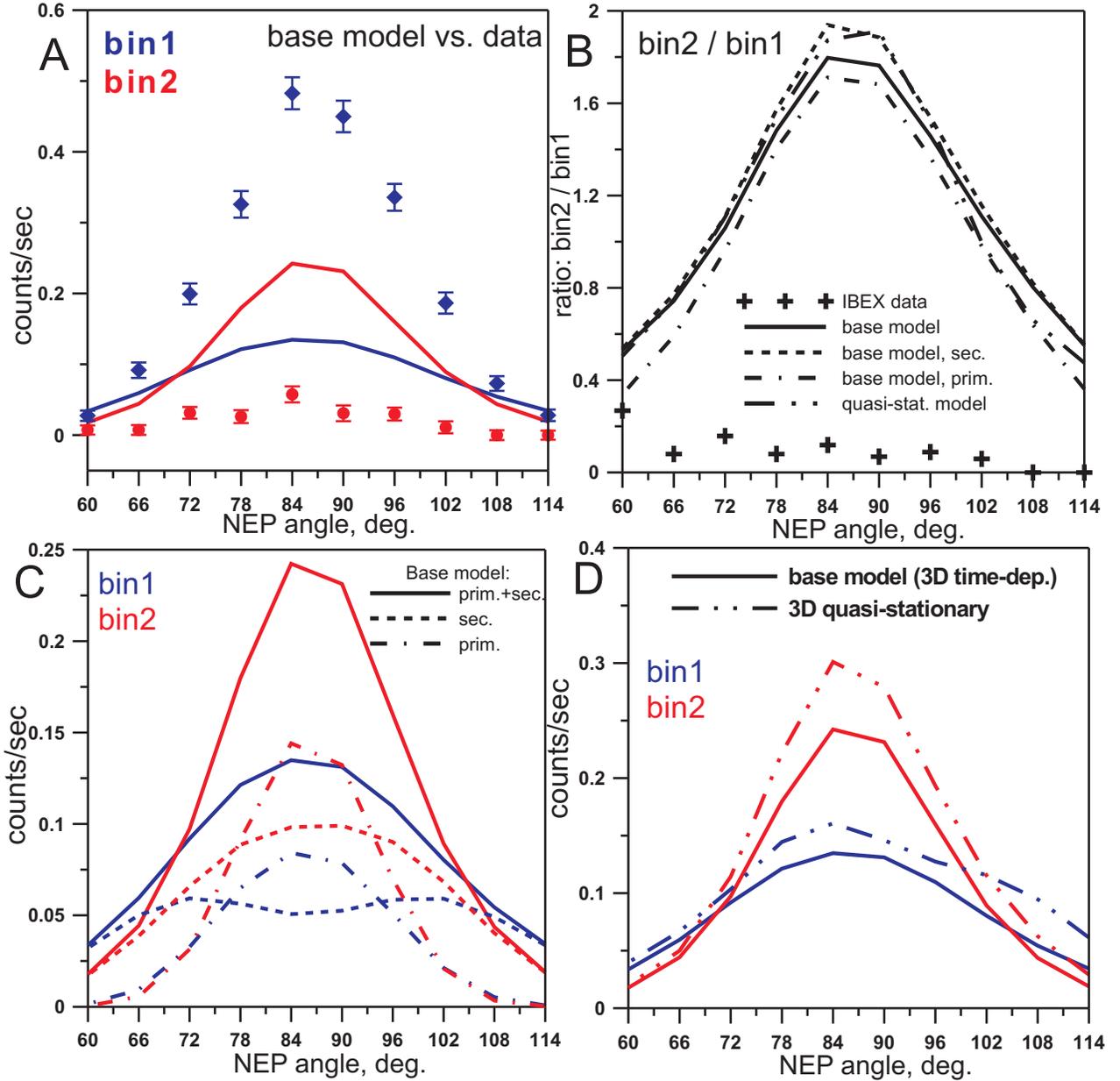}
\caption{Count rate of ISH atoms as functions of the NEP angle obtained by the numerical model for the geometry of the IBEX-Lo observations in energy bins 1 and 2 during orbit~23. A. Comparison between IBEX-Lo data and the results of the base 3D time-dependent kinetic model of the hydrogen distribution described in section~\ref{section_model}. B. Ratio of counts in energy bins 2 and 1 as functions of the NEP angle. The plot shows IBEX-Lo data (symbols), the results of the base model (solid curve), the results of the base model for the secondary population of hydrogen (dashed curve), the same but for the primary population (dashed-dotted curve), the results of the 3D quasi-stationary model (dashed-dotted-dotted curve).
C. The results of calculations in the frame of the base model separately for the primary and secondary populations.
D. Comparison between time-dependent and quasi-stationary models.
}\label{flux_3dtd}
\end{figure}

Fig.~\ref{flux_3dtd}~A shows the comparison of the data with the base model results.
It can be seen that there is qualitative difference between the data and model: the data shows that the count rate in energy bin~1 is much larger than that in energy bin~2, while
our state-of-the-art model provides a larger count rate in energy bin~2 (see also the solid curve in Fig.~\ref{flux_3dtd}~B for the ratio of count rates in bins 2 and 1;
other curves in this plot will be discussed below as well as plots C and D).

In principle, the obtained qualitative differences between the data and the base model may have two causes: 1) there are some problems with the model (e.g. lack of knowledge
of the model parameters or physical processes), 2) there are some inaccuracies in the processing of the IBEX-Lo data and/or the determination of the instrumental parameters
(e.g. geometrical factors, energy response functions, boundaries of energy bins, etc.). In this paper, we focus on the first possibility and analyze how the considered ISH fluxes depend on the parameters of our model. An investigation of the possible instrumental effects and how they influence the measured counts is proposed for future papers.

Generally, there are two subsets of model parameters. The first subset is the solar parameters, which determine the interaction between H atoms and the solar interior (photons and protons).
Namely, these parameters are $\mu_0$, $\gamma$, and $\beta_E$. They can be determined based on different observations of the Sun (measurements
of the solar radiation and the SW), but some uncertainties of these parameters may still be present. The second set of the model parameters is related to the boundary conditions for the ISH velocity distribution function taken at 90~AU from the Sun. As mentioned before, these boundary conditions are based on the results of the global kinetic-MHD model of the heliospheric interface.

In the following sections, we study how both sets of the model parameters affect the ratio of the count rate of
energy bins~2 and 1.

\subsection{Role of primary and secondary populations}

As mentioned previously, inside the heliosphere there are two populations of the interstellar hydrogen atoms: the primary (entered to the heliosphere without charge exchange) and the secondary (created by charge exchange in the heliospheric interface) populations. The properties of the primary and secondary populations are different. Namely,
 the secondary atoms have a smaller bulk velocity and larger temperature compared with the primary atoms. Therefore, it is interesting
  to study which population dominates in the considered IBEX-Lo energy channels. To answer this question, we performed corresponding calculations in the frame of the base 3D time-dependent model
separately for the primary and secondary interstellar atoms. The results are shown in Fig.~\ref{flux_3dtd}~C.
It is seen that for the NEP angle $\in[70^{\circ},100^{\circ}]$ the primary atoms dominate in both energy bins, while for the NEP angle at the flanks, on
the contrary, the secondary atoms dominate. However, Fig.~\ref{flux_3dtd}~B (dashed and dash-dotted curves) shows that the count ratio in the two energy bins is about
the same for the primary and secondary interstellar atoms and their mixture. This means that if we change the proportion between the
primary and secondary atoms in our model \citep[which is possible, e.g. by changing of the LISM parameters; see][]{izmod_etal_1999, izmod_2007} it will not help to resolve the qualitative discrepancy between the theoretical results and IBEX data.

\subsection{Investigation of the role of time-dependent effects}

Solar parameters vary significantly within a cycle of the solar activity. Therefore, before performing a parametric study for different magnitudes of
$\mu_0$, $\gamma$, and $\beta_E$ we need to analyze the role of time-dependent effects. Fig.~\ref{flux_3dtd}~D presents the comparison between the results
of the base 3D time-dependent model and the simplified 3D quasi-stationary model. In the latter case, the parameters $\mu(w_r)$ and $\beta_E(\lambda)$ do not depend on time
and correspond to their local values during the considered period of time. It can be seen from the figure that in the stationary case, the counts in both energy bins
are larger by about 20~\% than in the time-dependent case. This is due to the local
minimum in solar activity (i.e. previously the solar radiation pressure and ionization rate are higher) that occurred during IBEX orbit 23. Therefore, in the time-dependent case when previous periods of time are taken into account, a smaller number of
H atoms can reach the vicinity of the Sun. However, the general behavior of the count rates in the first and
second energy bins and their ratio is about the same for the stationary and non-stationary cases (compare the solid and dashed-dotted-dotted curves in Fig.~\ref{flux_3dtd}~B).
Therefore, time-dependent effects may be important for the analysis of the count rate, but they are not important when investigating
the qualitative difference between the model results and the IBEX-Lo data. This conclusion is consistent with the results of \citet{bzowski_rucinski_1995} and \citet{bzowski_etal_1997}, who studied the role of time-dependent effects and showed that the time delay between the local maximum of the solar radiation pressure and
the corresponding local minimum of hydrogen number density at 1~AU is almost zero.


\subsection{Calculations with different LISM parameters}

The distribution of the ISH at 90~AU which is used as the boundary conditions in our model is, on the one hand, the lesser known parameter of the model.
However, on the other hand, we can not choose it randomly because this distribution should be consistent with the global model of the SW/LISM interaction, and
we have many restrictions for the parameters of the global model based on experimental data from different spacecraft.
These restrictions are described by Izmodenov \& Alexashov (2015). In that companion paper, it is also shown that the kinetic-MHD model of the heliospheric
interface which we use here is consistent with much of the experimental data (although not all of them).

\begin{figure}
\includegraphics[scale=0.8]{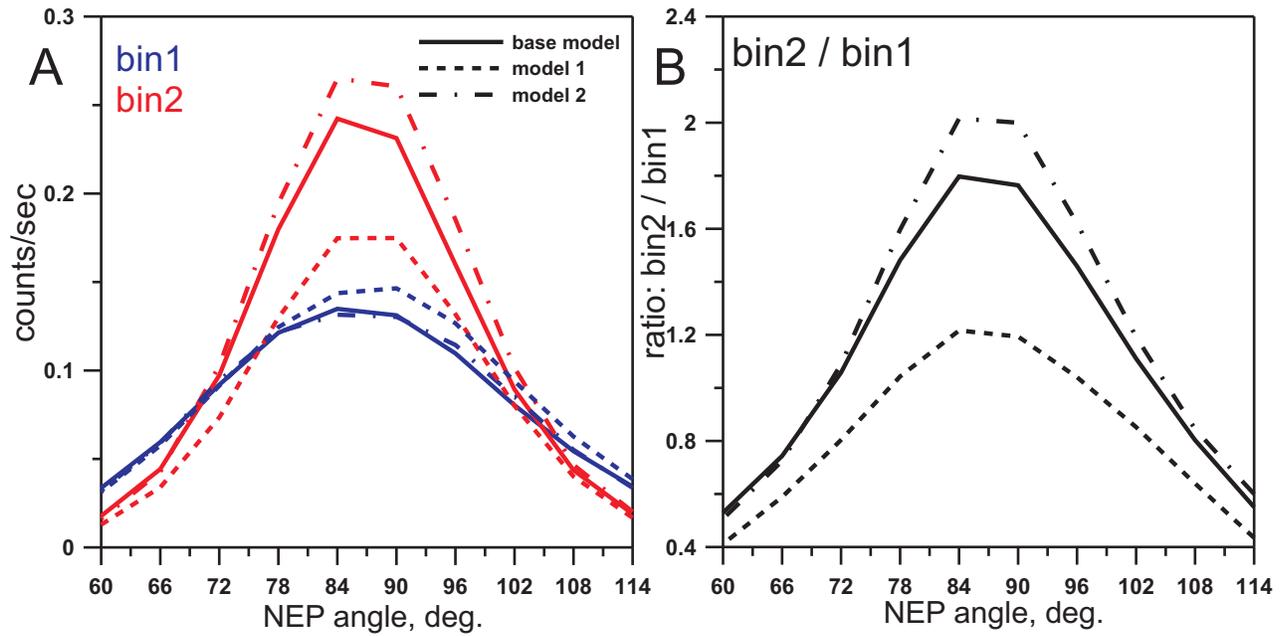}
\caption{ A. Count rate for energy bins 1 and 2 calculated for orbit 23 in the context of the base model and models with different LISM parameters (for a description
of models~1 and 2, see the text).
B. Ratio of counts in energy bins 2 and 1 (the types of curves are the same as in plot~(A)).
}\label{dif_LISM}
\end{figure}

Due to computational restrictions, we are not able to perform a full parametric study for different LISM parameters (because it requires numerous calculations
in the context of the global kinetic-MHD model of the heliospheric interface). To estimate the possible effect of
the applied LISM parameters, we perform two additional calculations using the following boundary conditions in the LISM (corresponding to recent results of  measurements of the interstellar helium fluxes):
\begin{itemize}
  \item Model~1: parameters are the same as in the base model (see section~2), except for the velocity vector $\textbf{V}_{LISM}$, which is taken from the results of the primary analysis of
  the IBEX-Lo helium data \citep{bzowski_etal_2012, mccomas_etal_2012, moebius_etal_2012}. Here, $V_{LISM}$=23.2~km/s, the ecliptic longitude (J2000) is 79$^{\circ}$,
  and the ecliptic latitude is -4.98$^{\circ}$. Note that this vector contradicts to the Ulysses helium data \citep{witte_2004, bzowski_etal_2014}.
  \item Model~2: parameters are the same as in the base model, except for the temperature $T_{LISM}$, which is increased to 8000 K. Such an increase is consistent with recent results obtained by several authors from reanalysis of the Ulysses/GAS and IBEX-Lo helium observations
      \citep{bzowski_etal_2014, katush_etal_2014, mccomas_etal_2015He, wood_etal_2015}.
\end{itemize}
The results of our calculations are shown in Fig.~\ref{dif_LISM}. It can be seen that increasing of the LISM temperature
leads to a small increase of the fluxes in the second energy bin compared to the base model (which is obvious because the atoms became to be more energetic). Changing of $\textbf{V}_{LISM}$ leads to decreased fluxes in energy bin~2 and a small increase in energy bin~1. This is caused by the decrease of the atoms' bulk velocity (more of them appeared in the first
energy bin) and also by the increased ionization loss of atoms with smaller velocity (the so-called selection effect).
The ratio of the count rates in the two energy bins is qualitatively the same for all of the models (see Fig.~\ref{dif_LISM}~B). Although, for the model~1 the ratio is a little bit
closer to the data than for other models, it is still greater than one, contrary to the data, and the absolute values
of the count rates for both bins is significantly different from the data.
 Therefore, acceptable changes of the
LISM parameters
do not allow us to resolve the qualitative contradictions between the model and the data.



\section{Results of the stationary model}

In this section, we perform calculations using the stationary version of our model with fixed values of $\mu_0$ and $\beta_{E,0}$.
Before studying how the ISH fluxes depend on these parameters, it is worthwhile to compare the results of our stationary model
with the standard hot model, which is commonly used for interpretations of different experimental data on neutrals in the heliosphere.
As we mentioned in the Introduction, the classical hot model assumes a one-component uniform Maxwellian distribution of interstellar hydrogen (a mixture of primary
and secondary) far away from the Sun (e.g. at 90~AU), while in our base model a 3D normal distribution with an angular dependence of the parameters at the boundary
sphere is assumed for the primary and secondary atoms separately.

Here, we performed calculations using the stationary model with constant values of $\mu=\mu_0=0.89$ and $\beta_E=4.63\cdot10^{-7}$~s$^{-1}$
(these values are taken from the non-stationary model at the considered time period).
Solar radiation pressure is assumed to be constant for all velocities, and we do not apply any heliolatitudinal variations of $\mu$ and $\beta$
(for comparison with the simple hot model).
In the case of our base model, the boundary conditions at 90~AU are taken to be the same as described in section~2, while for the hot model a simple
Maxwellian distribution is assumed. The parameters of this distribution are as follow:
number density $n_{mix}=0.094$~cm$^{-3}$, averaged velocity $V_{z,mix}$=-21.12~km/s ($V_x=V_y=0$), and averaged temperature $T_{av,mix}$=13962~K. These values
are kept the same at 90~AU (without angular dependence) and are taken from the results of the global heliospheric model for a mixture of primary and secondary atoms at
90 AU in the direction where most of the H atoms measured by IBEX come from.

\begin{figure}
\includegraphics[scale=0.8]{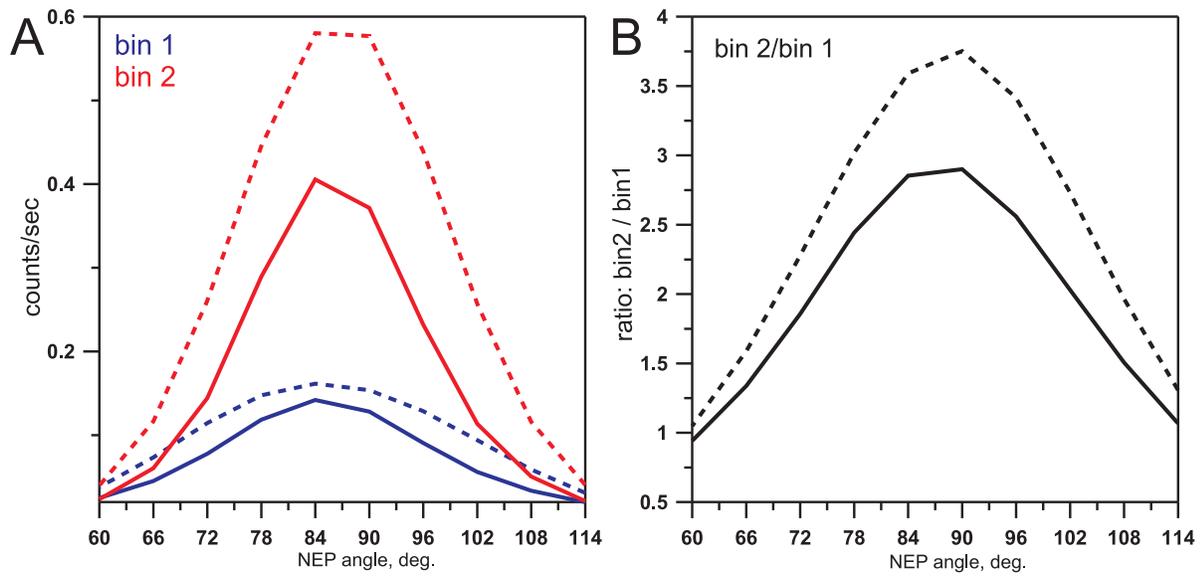}
\caption{Comparison between the base model (solid curves) and the standard hot model (dashed curves) with a uniform Maxwellian ISH distribution far away from the Sun. Results of the stationary models with constant $\mu=\mu_0=0.89$ and $\beta_E=4.63\cdot10^{-7}$~s$^{-1}$.
Count rate in orbit~23 for energy bin~1 and bin~2 (plot (A)) and their ratio (plot (B)).
}\label{maxw}
\end{figure}

Fig.~\ref{maxw} presents the results of our calculations.
It can be seen that the standard hot model leads to an overestimate of the counts in both energy bins compared with our model. This overestimate is caused by the fact that
the number density is kept constant across the whole boundary sphere in the hot model, but it decreases from the upwind to downwind in our model.
Also, the hot model gives a significantly larger ratio of counts in energy bins 2 and 1 than in our model.

This comparison shows that the hydrogen distribution assumed far away from the Sun is important for the ISH fluxes measured by IBEX-Lo and using
a simplified hot model may lead to incorrect interpretation of the data.

\subsection{Influence of the hydrogen ionization rate}

\begin{figure}
\includegraphics[scale=0.7]{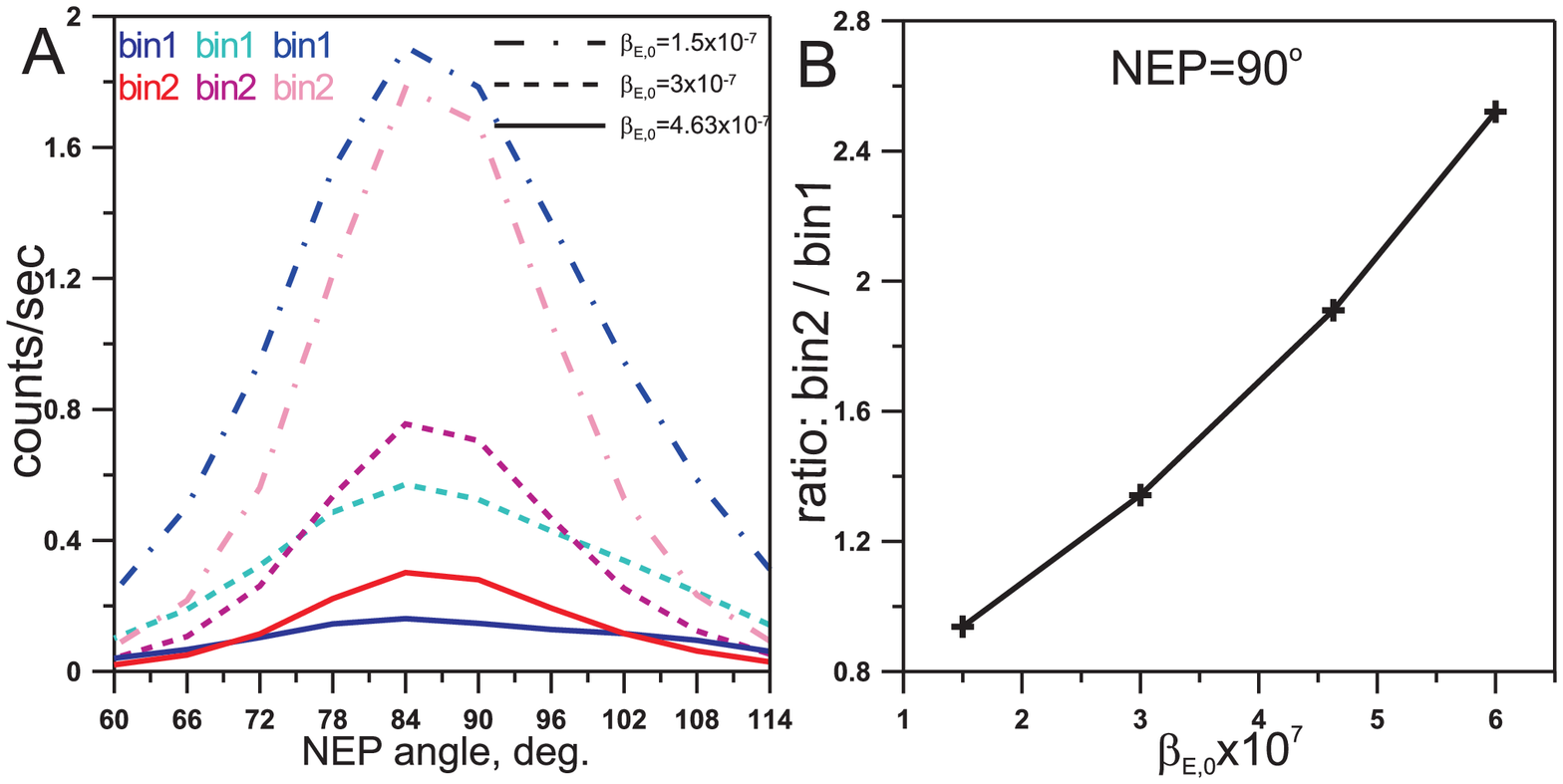}
\caption{Investigation of the role of the hydrogen ionization rate at 1~AU. All calculations are performed using the 3D quasi-stationary model with fixed
magnitudes of $\mu_0=0.89$ and $\gamma=0$. A. Count rate for energy bins 1 and 2 as a function of NEP angle. Results of the models with different values
of $\beta_{E,0}$: $\beta_{E,0}=4.63\cdot10^{-7}$~s$^{-1}$ (solid curve), $\beta_{E,0}=3\cdot10^{-7}$~s$^{-1}$ (dashed curve),
$\beta_{E,0}=1.5\cdot10^{-7}$~s$^{-1}$ (dashed-dotted curve).
}\label{flux_betta}
\end{figure}

In this section, we study the influence of the hydrogen ionization rate in the ecliptic plane ($\beta_{0,E}$) on the ISH fluxes measured by IBEX.
Calculations are performed using a 3D stationary version of the base model with $\mu_0=0.89$ and $\gamma=0$.
Fig.~\ref{flux_betta}~A presents the results of calculations with different values of $\beta_{0,E}$.
Fig.~\ref{flux_betta}~B presents the ratio of the counts in energy bins 2 and 1 as a function of $\beta_{0,E}$.
It is seen that an increase of $\beta_{0,E}$ leads to a decrease of the counts in both energy bins (higher ionization causes increased atomic loss), and to
 a monotonic increase of the ratio. The last result is caused by the kinetic selection effect: namely, larger ionization rates
lead to an increased loss of slow atoms (because they have more time to be ionized than faster atoms) and as a result the fraction of atoms in energy bin~2
relative to energy bin~1 increases. However, we see that even for $\beta_{0,E}=1.5\cdot10^{-7}$~s$^{-1}$ (which is extremely small), the ratio of
the count rates in bin~2 to bin~1 is equal to 0.9, which is much larger than the IBEX observed ratio of 0.1. This means that although the results depend
on the ionization rates, any reasonable changes of that cannot explain this large discrepancy between the model (with a realistic $\mu<1$) and the IBEX data.

%

\subsection{Dependence of fluxes on parameter $\mu$}

Here, we investigate the influence of the radiation pressure parametrized by $\mu_0$ and $\gamma$ (see equation~(3)).
The results of this subsection were obtained using the base model with fixed values of $\mu_0$ and $\gamma$.
First, the magnitude of $\mu_0=0.96$ (close to 1) has been fixed while the parameter $\gamma$ has been allowed to vary.

\begin{figure}
\includegraphics[scale=0.8]{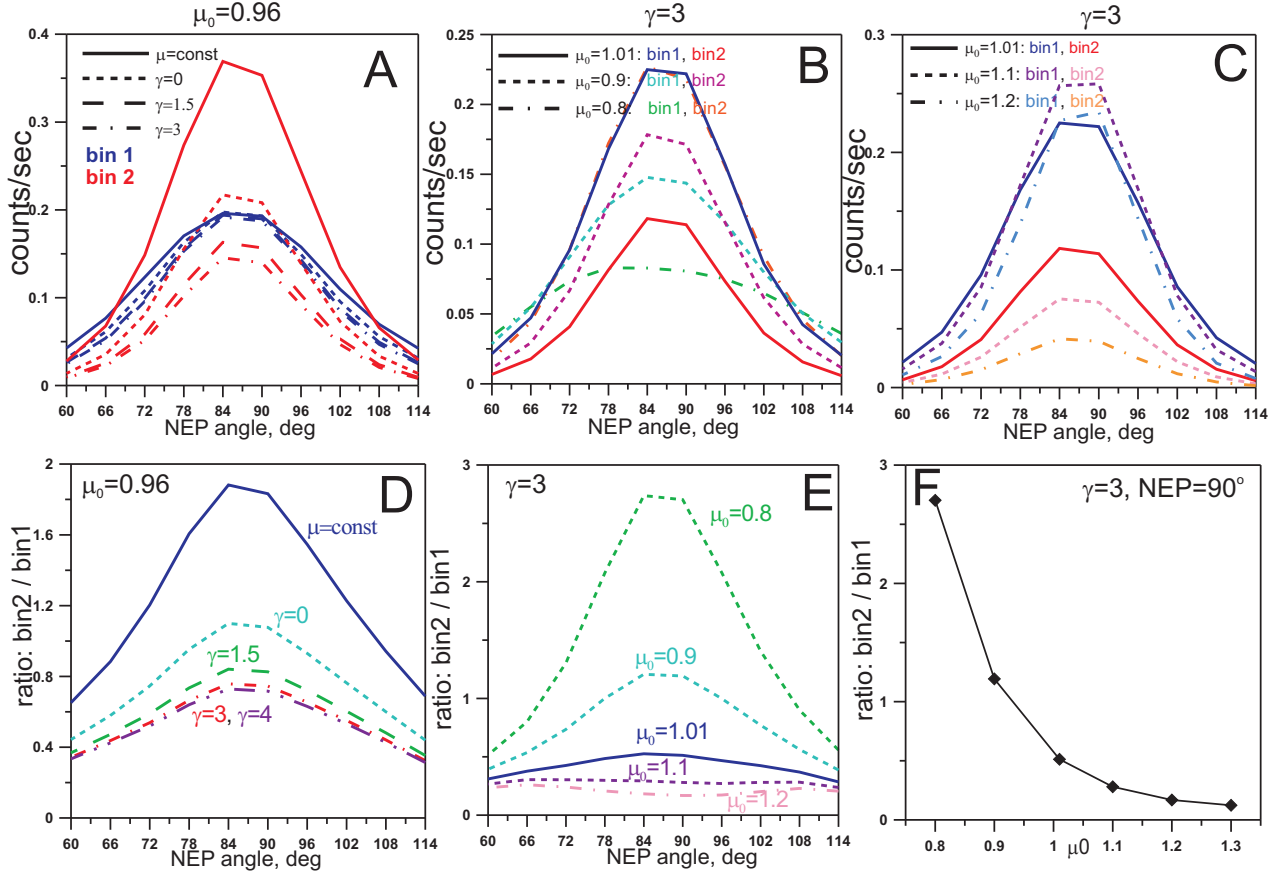}
\caption{Count rate of ISH atoms as functions of NEP angle obtained by the numerical model for the geometry of IBEX-Lo observations in energy bins 1 and 2 during orbit~23. A: Calculations for the different velocity dependence of $\mu$ and
a fixed value of $\mu_0=0.96$. Solid curve corresponds to constant $\mu$ (without dependence on velocity), while the other curves correspond to different $\gamma$ (see Fig.~\ref{ibex_geom}~A). B-C: Calculations for different values of
$\mu_0$ and fixed $\gamma=3$. D-E: Ratio of the count rate obtained in energy bin~2 to the count rate obtained in energy bin~1 for different $\gamma$ and $\mu_0$ corresponding to plots A-C. F: Ratio between the counts in bins 2 and 1
for $\alpha_{NEP}=90^{\circ}$ as a function of $\mu_0$ (for fixed $\gamma=3$).
}\label{H_flux_mu}
\end{figure}

Fig.~\ref{H_flux_mu}~A presents the results of our calculations. The count rate as a function of an NEP angle is shown for the energy bin~1 (blue curves)
and the energy bin~2 (red curves).
The changes of $\mu(v_r)$ mostly influence the count rate in energy bin~2 because
(as was mentioned before) the velocity dependence of $\mu$ is important for those atoms which speeds of more than 30-70~km/s,
which corresponds to a relative (to IBEX) velocity of $\sim$60-100~km/s or 19-52~eV. Such energies correspond to the second energy bin of the IBEX-Lo.
 From the plot (compare solid and all other red curves), we see that taking into account the velocity dependence of $\mu$ leads to a significant decrease of the counts in energy bin~2.
Fig.~\ref{H_flux_mu}~D presents the ratio between the counts in
energy bin~2 to the counts in energy bin~1. For $\alpha_{NEP}=90^{\circ}$, the ratio changes by more than 2.5 times for the cases
with $\mu$=const and $\gamma=3$. Therefore, our results show that taking into account the velocity dependence
of $\mu$ (which describes the self-reversal of the solar Lyman-alpha line) is extremely important for the ratio of the count rates measured in energy bins 2 and 1.
We also see from the plot that, as expected,
variations of $\gamma$ are not very important for the results (especially for $\gamma>1.5$) due to the weak dependence of $\mu$ on $\gamma$
 (see Fig.~\ref{mu_t_vr}~B).
 Note that the role of velocity-dependent solar radiation pressure on the interstellar hydrogen parameters
near the Sun was studied by \citet{tarnopolsky_bzowski_2009}. They found that the main difference between model results with and without velocity dependence of $\mu$
is a factor of 1.5 for the hydrogen distribution at 1~AU. Therefore, our results are consistent with these previous studies.

We also performed calculations with fixed $\gamma=3$ and different values of $\mu_0$ (Fig.~\ref{H_flux_mu}~(B) and (C) for counts and (E) and (F) for ratio). The increase of $\mu_0$ leads to a
decrease of the count rates for both energy bins (one exception for bin~1 and $\mu_0=1.01$ and $\mu_0=1.1$ demonstrates that the effect is not monotonic),
 and to a monotonic decrease of the ratio of the count rates in bins~2 and bin~1 (plot~F in Fig.~\ref{H_flux_mu}).
 Note that IBEX-Lo data have a bin~2 to bin~1 ratio of about 0.1, while the base time-dependent model predicts 1.7. Obviously, an increase of $\mu_0$ can
 resolve this problem.

The decrease in the count rates for both energy bins with increasing $\mu_0$ is because larger $\mu_0$ (i.e. larger solar
radiation force) leads to the deceleration of H atoms and the deflection of their trajectories. Therefore, fewer H atoms can reach Earth's orbit. The decrease of the ratio of bin~2 to bin~1 implies that
the fluxes of H atoms in bin~2 decrease more rapidly than in bin~1.
This is because the more energetic atoms in energy bin~2 are strongly affected by the velocity dependence of the solar radiation pressure.
Therefore, these atoms are more strongly deflected from the Sun than the slower atoms in energy bin~1.
Fig.~\ref{H_flux_mu}~F shows that variations of $\mu_0$ from 0.8 to 1.3 (a factor of 1.6) lead to enormous
changes in the ratio of the energy bin~2 counts to the energy bin~1 counts: this ratio decreases from 2.7 to 0.13, i.e. by more than 20 times.

Thus, the IBEX-Lo ISH data and, particularly, the ratio between the count rates measured in the first and second energy bins is very sensitive to the solar radiation pressure. Also, our parametric study shows that only an increase of the parameters $\mu_0$ and $\gamma$ can sufficiently decrease
the bin~2 to bin~1 ratio to reach a
qualitative agreement with the IBEX-Lo data.

\section{Fitting of the data for orbit~23 in 2009}\label{section_fitting}

In this section, we fit model parameters to the IBEX-Lo data (for orbit~23).
Test calculations show that with reasonable parameter choices, the results of the time-dependent model (with
hydrogen ionization rates taken from experimental data) cannot be made to fit well to the IBEX-Lo data. Therefore, we perform a fitting procedure using
 the 3D stationary model of the ISH distribution in the heliosphere. Temporal variations of the ionization rate are not included
in the model, and so the total ionization rates at 1~AU depend only on heliolatitude ($\beta_E(\lambda)$).
 Therefore, the search for the best-fit solution is performed by varying three parameters: $\mu_0$, $\gamma$, and $\beta_{E,0}$, where the last
 parameter is the total ionization rate of H atoms at 1~AU and zero heliolatitude. These parameters are determined by the least-square method through the minimization of $\chi^2$, defined as:
 \[
   \chi^2(\textbf{a})=\frac{1}{N-M}\sum_{i=1}^2 \, \sum_{j=1}^{10} \left( \frac{C_{i,j}(\textbf{a})-C_{i,j}^{data}}{\sigma_{i,j}^{data}}  \right)^2,
 \]
where $\textbf{a}$ is a vector of the three free parameters ($M$=3), $j$ is the index of the summation over 10 lines of sight ($N=2\times10$ is the total number
of used experimental data points for the two energy bins),
$C_{i,j}(\textbf{a})$ are the count rates calculated in the model for fixed set of free parameters, $C_{i,j}^{data}$ are
count rates obtained from IBEX, and $(\sigma_{i,j}^{data})^2$ is the variance associated with the measured count
rates.

\begin{figure}
\includegraphics[scale=0.8]{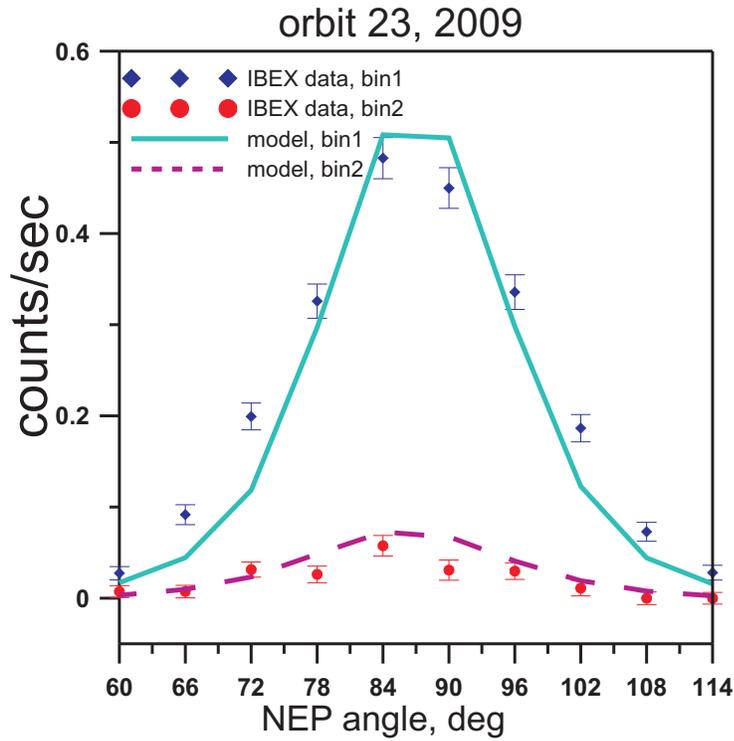}
\caption{Comparison between the IBEX-Lo data (orbit 23) and the model with the obtained best-fit parameters ($\mu_0=1.26$, $\gamma=3.5$, $\beta_{E,0}=3.7\cdot 10^{-7}$~$s^{-1}$ and for this parameter's set $\chi^2=6.82$). Results are presented
for energy bins 1 and 2. Error bars are shown for the data.
}\label{H_flux_best_fit}
\end{figure}

As a result of the $\chi^2$ minimization, the following parameters are found: $\mu_0=1.26^{+0.06}_{-0.076}$, $\beta_{E,0}=3.7^{+0.39}_{-0.35}\times 10^{-7}$~s$^{-1}$, and $\gamma=3.5^{+?}_{-3.02}$. The upper bound for $\gamma$ cannot be
determined because results are not sensitive to the magnitude of $\gamma$ for any value of $\gamma>0.48$. For the best-fit parameter set,
 we found $\chi^2_{min}=6.82$. Analyses of the obtained $\chi^2$ values and the procedure for calculating
 the uncertainties are presented in Appendix~B.
A comparison between the IBEX-Lo data and the model results for the best-fit solution is presented in Fig.~\ref{H_flux_best_fit}. There is a quite good agreement between the data and the model, although it seems that the data show a somewhat wider distribution for bin~1 compared to the model results.

Note that the determined magnitude of the total ionization rate ($3.7\cdot 10^{-7}$~$s^{-1}$) is about 20~\% smaller than the ionization rate known from measurements in the ecliptic plane (OMNI2 and SOLAR2000 databases), which yield $4.63\cdot 10^{-7}$~$s^{-1}$ for the period of orbit 23. The obtained value of $\mu_0=1.26$
is significantly larger than expected from observations (from measurements of the integrated Lyman-alpha flux transformed
 to the flux at the line center we get $\mu_0=0.89$, see Fig.~\ref{mu_t_vr}~A).
We also note that if we fix $\mu_0=0.89$ and try to fit the model parameters to the data, we obtain $\chi^2\geq 90$ for any $\gamma\in[0,4]$ and
 $\beta_{E,0}\in[2\cdot10^{-7}, 6\cdot10^{-7}]$~s$^{-1}$. Hence, we are not able to find an appropriate solution for $\mu_0\leq1$.

\section{Summary and Discussion}

Analysis of the IBEX-Lo measurements of interstellar hydrogen during orbit 23 in 2009 has been performed using a state-of-the-art kinetic model of the ISH distribution
in the heliosphere. We show that the base 3D time-dependent version of the model leads to a qualitative disagreement between the IBEX data (the ratio of counts
in the first and second energy bins) and model simulations. We perform test calculations to study the influence of different model parameters on the ratio of the energy bin~2 to the energy bin~1 count rates.
We show that when using the appropriate models of the heliospheric interface consistent with different experimental data (without dramatic
changes in our concept of the heliosphere), only variations of the parameter $\mu$ allow us to obtain qualitative agreement between the theoretical results and the IBEX-Lo data.

We have studied the influence of the solar radiation pressure and its velocity dependence on the ISH count rate measured by IBEX-Lo during orbit~23. It is shown that the increase of
$\mu_0$ (i.e. the value that corresponds to $w_r=0$) from 0.8 to 1.3 results in a decrease of the counts in both energy bins~1 and 2 and a sharp decrease (from 2.7 to 0.13) of the ratio of bin~2 counts to bin~1. It is also shown that including the velocity dependence of
$\mu$ is very important. Modeling in which the dependence of $\mu$ on the radial velocity is not taken into account leads to considerable overestimation of the count rate in energy bin~2. Changes in the parameter $\gamma$ (which is responsible for the self-reversal shape of the solar Lyman-alpha profile) lead to variations of the count rates measured in the second energy bin for $\gamma \in [0, 1.5]$ but almost does not influence the results for $\gamma>1.5$.

Thus, both $\mu_0$ and $\gamma$ have a strong influence on the count rate in energy bin~2 and on
the ratio of bin~2 counts to bin~1 counts. Therefore, precise information on the solar radiation pressure is
critically important for models used to analyze and interpret the IBEX-Lo ISH data.

Fitting of the model to IBEX-Lo data (for orbit 23, end of March 2009) is performed using our 3D stationary kinetic model of the ISH distribution with H parameters at 90~AU taken from the global model of the heliospheric interface. We find that $\chi^2$ approaches its minimum for the following
set of the model parameters: $\mu_0=1.26^{+0.06}_{-0.076}$, $\gamma\geq0.48$, and a total hydrogen ionization rate at 1~AU of $\beta_{E,0}=3.7^{+0.39}_{-0.35}\times 10^{-7}$~s$^{-1}$. The obtained magnitude of $\mu_0$ is significantly larger than the value (0.89) derived from measurements of the solar Lyman-alpha irradiance for
the considered time period.

In general, there are three possible ways to account for the discovered discrepancies between inferred values of $\mu_0$:

1. It is possible that the local $\mu$ obtained from
measurements of the solar irradiance is underestimated. This could be caused by uncertainties in the absolute calibration of the instruments. \citet{lemaire_etal_2005} indicated that
uncertainties in the calibration factor of UARS/SOLSTICE and SOHO/SUMER are about $\pm10$~\%. Data from these instruments are used to obtain the integrated
solar Lyman-alpha flux and transform it to the flux at the line center (which in turn is needed to determine $\mu_0$). It is unlikely that $\mu$ can
be changed from 0.89 to 1.2-1.3 due to uncertainties in calibration alone, although our analysis of the IBEX-Lo data raises questions about the accuracy of our knowledge
of the absolute values of the solar Lyman-alpha flux.

2. It is possible that shortcomings of our model of the ISH distribution in the heliosphere lead to an overestimate of $\mu_0$.
For example, variations of shape of the solar Lyman-alpha spectrum (and the corresponding dependence of $\mu$ on $w_r$) with the solar cycle are ignored in the model, although these variations (not very large) are obtained from measurements by \citet{lemaire_etal_2002, lemaire_etal_2005}.
Also, our method for the reconstruction of the latitudinal variations of the hydrogen ionization rate (based on Lyman-alpha data) allows us to describe global variations of the SW with heliolatitude, but  misses some local variations that can be important for the IBEX-Lo measurements. It is likely that the time resolution
 in the model is insufficient (we used the data on $\mu(t, w_r)$ and $\beta_{E,0}(t)$ averaged over one Carrington rotation i.e. the time resolution is about 1 month, while for the IBEX-lo data very local values of $\mu$ can be important). Further detailed investigations of the role of local (short timescale) temporal and heliolatitudinal effects on the ISH fluxes measured by IBEX-Lo are needed to resolve this problem.
 Another possible problem of our model can be connected to the LISM parameters adopted in the model. We present calculations with three
 sets of the LISM parameters, which lead to close results. In our calculations, we use the results of a global kinetic-MHD self-consistent model of the heliospheric interface. Parametric studies using this model and
 many different sets of LISM parameters require large amounts of computational time, and therefore are beyond the scope of this study.
  Also, we have
 restrictions for the possible LISM parameters from other experimental data and cannot choose them arbitrarily.

3. The third contributing reason for the discrepancy in $\mu_0$ might be related to details of the instrumental response of IBEX-Lo (e.g., the energy-dependent geometric factors and response functions). Recently, \citet{fuselier_etal_2014} proposed that the uncertainties of the count rate in the first two bins of IBEX-Lo could be about 50~\%. Future work should continue to include the ever-increasing sophistication of the detailed instrumental response and decrease this uncertainty.

We should note that recently the distribution of the interstellar hydrogen inside 1~AU from the Sun was studied remotely by cross-analysis of the backscattered
Lyman-alpha intensities measured by SOHO/SWAN and MESSENGER/MASCS \citep{quemerais_etal_2014}. Our 3D time-dependent model of the hydrogen distribution in the heliosphere was applied to that analysis. It was found that the modelled hydrogen distribution obtained in 2009 (1-2 years before observations) provides good agreement with the data for the
Lyman-alpha intensity between SOHO and MESSENGER, but underestimates the distance to the MER. The conclusion reached is that the modeled hydrogen atoms are capable (on average) of moving  too close to the Sun when compared to the observations. In order to increase the distance to the MER in the model, we need to increase either the hydrogen ionization rate or the radiation pressure ($\mu_0$). The second scenario is consistent with our current results obtained for the IBEX-Lo data.

\acknowledgments

O.K., V.I., and D.A. were supported by RFBR grant No.~14-02-00746 for part of the analysis of the IBEX-Lo data. The supporting numerical modeling of the global heliosphere/astrosphere has been done in the frame of RSF grant 14-12-01096.
N.S. and D.M. were supported by the IBEX mission, which is part of NASA's
Explorer Program.
This work is done under discussions of international ISSI teams No.~318 and 327.
Calculations of the ISH distribution were performed by using the Supercomputing Center of Lomonosov Moscow State University (supercomputers ``Lomonosov'' and
``Chebyshev'').

\appendix

\section{Transformation of model fluxes to IBEX-Lo count rate}

\begin{figure}
\includegraphics[scale=0.8]{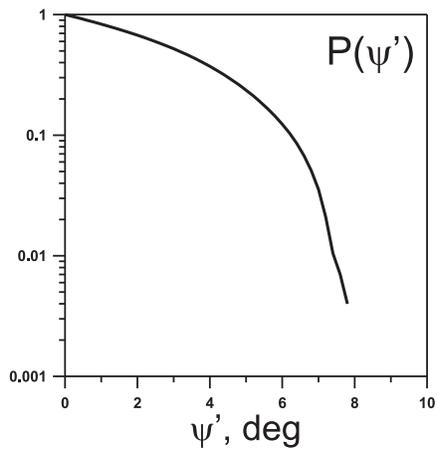}
\caption{Collimator transmission (or point-spread) function; $\psi'$ is
an angle from the axis of the collimator.
}\label{point-spread}
\end{figure}

For comparison with the IBEX-Lo data, one needs to convert the fluxes calculated in the model to the count rate (number of counts per second). To do this,
we need to integrate the fluxes
over a $6^{\circ}$ bin of IBEX's lines of sight, acceptance angles of the collimator, and the corresponding energy range.
The formula for the count rate in energy bin $i$ and for the NEP angle $\alpha_j$ is the following \citep[this is an analog of formula~3 from][]{schwadron_etal_2013}:
\begin{eqnarray}
C_{i,j}&=\frac{1}{\Delta t} \, \int_{t_0}^{t_1} \, dt \; \frac{1}{\Delta\alpha} \, \int_{\alpha_j-\Delta\alpha/2}^{\alpha_j+\Delta\alpha/2} \, d \alpha \quad \times \nonumber\\
&\times \int \int \hat{P}(\varphi',\psi') \, d\varphi' \, d\psi' \; \int_{V_{i,1}}^{V_{i,2}}
f_H(\textbf{w}_H) \, |w_{rel}|^3 \, E_{rel} \, G_i \, \hat{T}_i(E_{rel}) \, d w_{rel}.  \label{counts}
\end{eqnarray}
Here, $\Delta t = t_1-t_0$ is the duration of the observations (in seconds) and the NEP-angle $\alpha$ determines the direction
of a line of sight in the observational plane $\pi$.
This angle varies over the range of $[\alpha_j-\Delta\alpha/2, \alpha_j+\Delta\alpha/2]$ centered at $\alpha_j$ with angular bin-width $\Delta \alpha = 6^{\circ}$. Integration over the collimator is represented by a collimator
transmission function (sometimes it is called the point spread function) $\hat{P}(\varphi',\psi')$ which
determines the probability of atom's detection inside the collimator \citep[see][for details]{schwadron_etal_2009}.
In our calculations, we use a simplified conical shape of the collimator (instead of a realistic hexagonal shape) because
numerical tests show that this approximation is appropriate and does not influence the results.
In this case, $P$ depends only on one angle $\psi'$ counted from the axis of the collimator. We use $P(\psi')$
found from the ISOC datacenter (the plot is presented in Fig.~\ref{point-spread}).
In equation~(\ref{counts}), $f_H$ is the velocity distribution function of the ISH atoms at the point of observation, $w_{rel}$ is the atom velocity relative to the spacecraft, $\textbf{w}_H$
is the absolute atom's velocity vector (i.e. $\textbf{w}_H=\textbf{w}_{rel}+\textbf{V}_{SC}$, where $\textbf{V}_{SC}$ is the spacecraft velocity and the direction of $\textbf{w}_{rel}$ is determined by the local line of sight inside the collimator);
$E_{rel}=m_H\, w_{rel}^2/2$, $V_{i,1}$ and $V_{i,2}$ determine the boundaries of energy bin~i:
$E_{min,i}=m_H\, V_{i,1}^2/2$ and $E_{max,i}=m_H\, V_{i,2}^2/2$, $m_H$ is the mass of an H atom; the boundaries of the energy ranges for bin~1 and bin~2 are taken from \citet{schwadron_etal_2013} and listed in Table~1. $G_i$ is the geometrical factor
(constant for each energy-bin), magnitudes of $G_i$ for $i=1,2$ are also listed in Table~1. Let us emphasize that $G_2$ is larger than $G_1$ almost
by a factor of two. Hence, the same hydrogen fluxes in the two energy bins will give a two times larger count rate in energy bin~2 than in energy bin~1.
Integration over the energy bin is performed with the normalized energy transmission function $\hat{T}_i(E)$ taken
from \citet{schwadron_etal_2013}:
\begin{eqnarray}
 T_i(E)&=exp\left(-4 \, \ln2\, \frac{(E/E_{c,i}-1)^2}{\Delta_1^2}\right) \quad \mathrm{for} \quad E\leq E_{c,i} \nonumber\\
       &=exp\left(-4 \, \ln2\, \frac{(E_{c,i}/E-1)^2}{\Delta_2^2}\right) \quad \mathrm{for} \quad E>E_{c,i},
\end{eqnarray}
where $E_{c,i}$ is the central energy of a given energy bin (see Table~1), and $\Delta_1=2(1-E_{min,i}/E_{c,i})$,
$\Delta_2=2(1-E_{c,i}/E_{max,i})$.
Functions $\hat{P}$ and $\hat{T}$ in formula~\ref{counts} are normalized by the following:
\begin{equation}
 \hat{P}(\acute{\varphi},\acute{\psi})=\frac{P(\acute{\varphi},\acute{\psi})}{\int \int P(\acute{\varphi},\acute{\psi}) d\acute{\varphi} \, d\acute{\psi}}, \quad
 \hat{T_i}(E)=\frac{T_i(E)}{\int_{E_{min,i}}^{E_{max,i}} T_i(E) \, dE},
\end{equation}
where integrations are performed over the acceptance angles inside the collimator and the energy range, respectively.

\section{Analysis of $\chi^2$ and calculations of uncertainties}\label{section_chi2}

\begin{figure}
\includegraphics[scale=0.8]{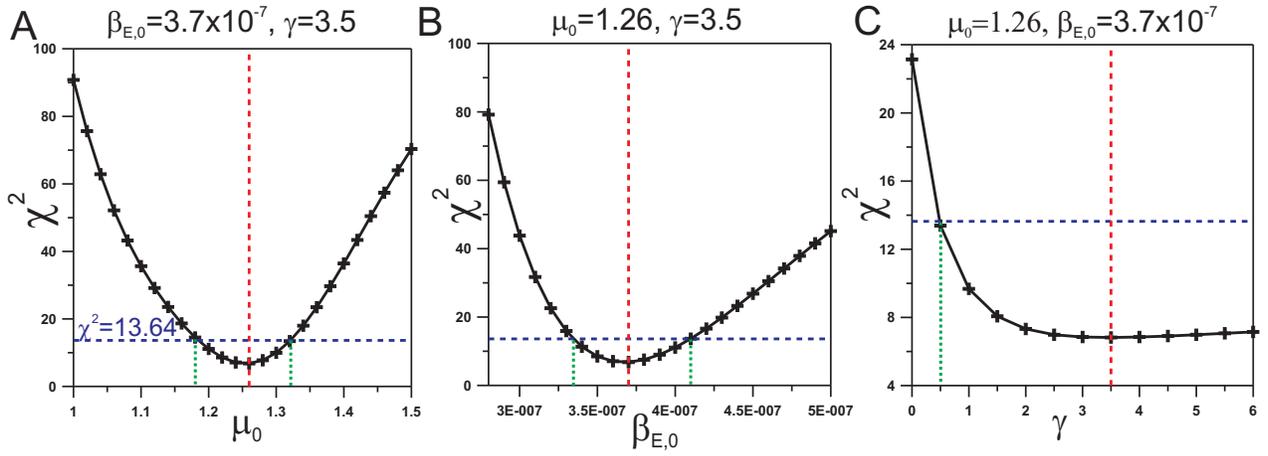}
\caption{Obtained $\chi^2$ in the fitting procedure as a function of $\mu_0$ (A), $\beta_{E,0}$ (B) and $\gamma$ (C). Red vertical lines in each plot
correspond to the minimum of $\chi^2$. Blue horizontal lines show $\chi^2=13.64$ that is found as a level of error bars (see section~\ref{section_chi2}).
Green dotted lines show corresponding ranges of the model parameters (for $\chi^2\leq13.64$).
}\label{chi2}
\end{figure}

Fig.~\ref{chi2} shows the obtained $\chi^2$ as a function of the parameters $\mu_0$, $\beta_{E,0}$ and $\gamma$. For each plot, two of the three parameters are fixed
and correspond to the determined best-fit magnitudes and the third parameter is varied. We see that for $\mu_0$ and $\beta_{E,0}$ the minimum of $\chi^2$
is quite deep, while for $\gamma$ the minimum almost disappears ($\chi^2$ is almost constant for $\gamma>1.5$). Therefore, $\gamma$ cannot be determined
precisely from the fitting of the data and only a lower limit of $\gamma$ can be provided.

The standard method for calculations of uncertainties for the determined best-fit parameters in the least-square method is to take
$\chi^2_0=\chi^2_{min}+1$ and find the range of parameters corresponding to $\chi^2\leq \chi^2_0$. However, this procedure is valid if
the $\chi^2_{min}$ obtained is close to 1. This is not our case because we found $\chi^2_{min}=6.82$. Theoretically, this means that either
IBEX data uncertainties ($\sigma_{i,j}^{data}$) are underestimated, or that we need to add some uncertainty connected to our numerical model.
 We introduce artificial model uncertainties $\sigma_{i,j}^m=\alpha\cdot \sigma_{i,j}^{data}$ such that the minimum $\chi_1^2$ would be equal to 1, i.e.,
\[
 \chi_1^2(\textbf{a})=\frac{1}{N-M}\sum_{i=1}^2 \, \sum_{j=1}^{10}  \frac{(C_{i,j}(\textbf{a})-C_{i,j}^{data})^2}{(\sigma_{i,j}^{data})^2\cdot(1+\alpha^2)}  = \frac{1}{1+\alpha^2}\cdot \chi^2(\textbf{a}),
\]
and $\alpha$ is chosen such that
\[
 1=\chi_{1,min}=\frac{1}{1+\alpha^2}\cdot \chi^2_{min}.
\]
Therefore, $1+\alpha^2=\chi^2_{min}=6.82$. Next, we consider the condition $\chi^2_1<\chi^2_{1,min}+1=2$, which gives $\chi^2<2(1+\alpha^2)=2\cdot \chi^2_{min}=13.64$.
From this condition and plots A-B in Fig.~\ref{chi2} we can find uncertainties for the obtained best-fit parameters. Namely,
$\mu_0=1.26^{+0.06}_{-0.076}$, $\beta_{E,0}=3.7^{+0.39}_{-0.35}\times 10^{-7}$~s$^{-1}$, $\gamma=3.5^{+?}_{-3.02}$. The upper bound for $\gamma$ can not be
determined because the results are not sensitive to the magnitude of $\gamma$ for any $\gamma>0.5$.

\clearpage

\clearpage

\clearpage

\clearpage

\clearpage

\clearpage

\clearpage

\clearpage

\clearpage

\clearpage

\clearpage

\begin{table}
\begin{center}
\caption{Central energies ($E_c$), energy ranges ($E_{min}$ and $E_{max}$) and geometrical factors ($G$) for energy bin~1 and energy bin~2 of the IBEX-Lo sensor.\label{table_1}}
\begin{tabular}{ccccc}
\tableline\tableline
energy bin & $E_c$, eV & $E_{min}$, eV & $E_{max}$, eV & G, $cm^2 \, sr \, keV/keV$ \\
\tableline
1 &15 &11 &21  &$7.29\cdot10^{-6}$ \\
2 &29 &20 &41 &$1.41\cdot10^{-5}$ \\
\tableline
\end{tabular}
\end{center}
\end{table}

\end{document}